\documentclass[aps,prl,twocolumn,superscriptaddress,showpacs,floatfix,10pt]{revtex4-1}

\usepackage{amsmath}
\usepackage{graphicx}
\usepackage{color}
\usepackage{epstopdf}
\usepackage{natbib}
\usepackage{xfrac}
\usepackage{wasysym}
\usepackage{multirow}

\usepackage[margin=10pt,format=hang,justification=raggedright,singlelinecheck=false,caption=false]{subfig}

\usepackage{adjustbox} 

\usepackage{xcolor,colortbl}
\definecolor{Gray}{gray}{0.85}
\definecolor{lblue}{rgb}{0.8,0.8,1}  
\definecolor{blue}{rgb}{0.6,0.6,0.9}  
\usepackage[normalem]{ulem}

\newcommand{\eref}[1]{Eq.~(\ref{#1})}
\newcommand{\fref}[1]{Fig.~\ref{#1}}

\newcommand{\svek}{\mathbf}

\newcommand{\etal}{{\it et al.}}

\renewcommand{\Im}{\hbox{Im}}
\renewcommand{\Re}{\hbox{Re}}

\begin{document}

\title{Resistivity saturation in Kondo insulators}

\author{M.~Pickem}
\affiliation{Institute for Solid State Physics, TU Wien,  Vienna, Austria}
\author{E.~Maggio}
\affiliation{Institute for Solid State Physics, TU Wien,  Vienna, Austria}
\author{Jan M.~Tomczak}
\email{tomczak.jm@gmail.com}
\affiliation{Institute for Solid State Physics, TU Wien,  Vienna, Austria}

\date{\today}

\begin{abstract}
Resistivities of heavy-fermion insulators typically saturate below a characteristic 
temperature $T^*$.
For some, metallic surface states, potentially from a non-trivial bulk topology, are
a likely source of residual conduction.
Here, we establish an alternative mechanism:
At low temperature, in addition to the charge gap, the scattering rate turns into a relevant energy scale, invalidating the semiclassical Boltzmann picture. Finite lifetimes of intrinsic carriers limit conduction, impose the existence of a crossover $T^*$, and control---now on par with the gap---the quantum regime emerging below it.
We showcase the mechanism with realistic many-body simulations and elucidate how the saturation regime of the Kondo insulator Ce$_3$Bi$_4$Pt$_3$, for which residual conduction is a bulk property, evolves under external pressure and varying disorder.
Using a phenomenological formula we derived for the quantum regime, we also unriddle the ill-understood bulk conductivity of SmB$_6$---demonstrating that our mechanism is widely applicable to correlated narrow-gap semiconductors.
\end{abstract}

\maketitle

\begin{figure}[!th]
    {\includegraphics[width=\linewidth]{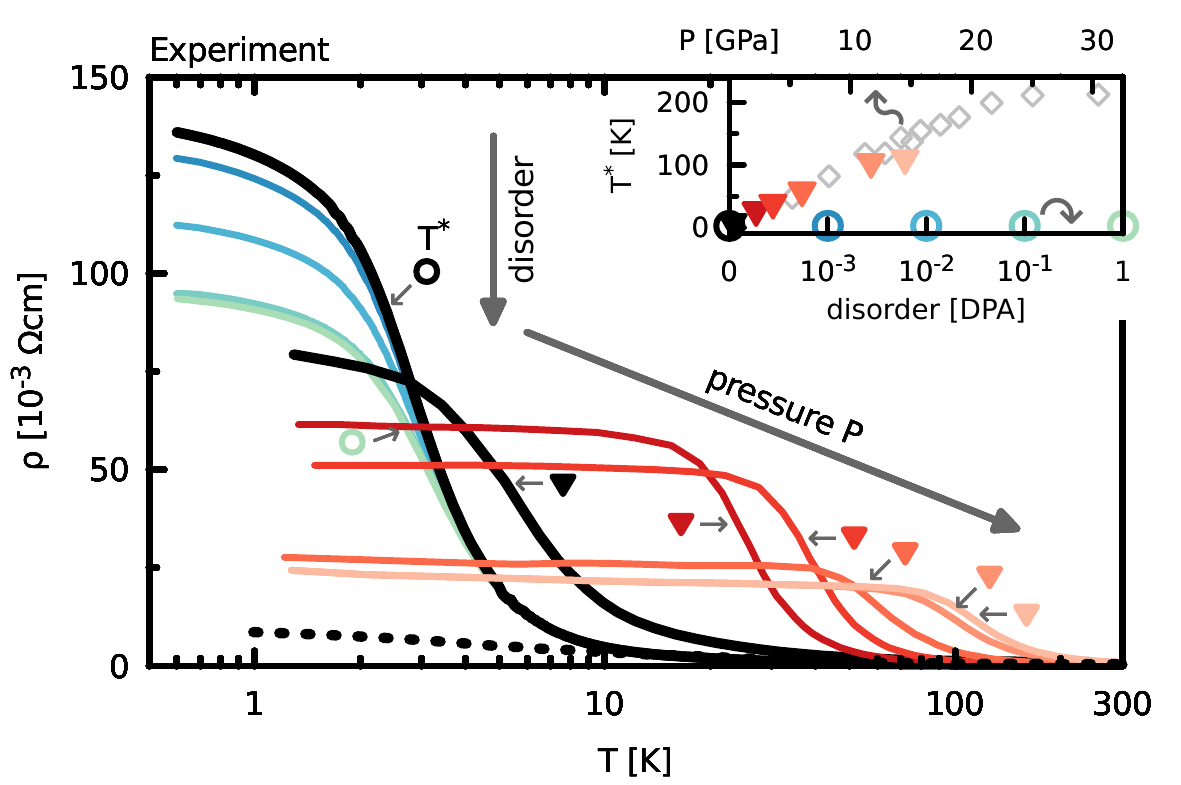}} 
\caption{{\bf Resistivity saturation in Ce$_3$Bi$_4$Pt$_3$.} 
Below an inflection temperature $T^*$ (indicated by small arrows), experimental resistivities $\rho(T)$\cite{PhysRevB.55.7533,PhysRevB.94.035127,PhysRevB.100.235133} deviate from activation-like behaviour ($T>T^*$) 
and enter a regime of {\it resistivity saturation} ($T<T^*$)%
---the focus of this work.
Pressure and disorder affect the resistivity differently:
Under pressure (black and shades of red; from Cooley \etal\cite{PhysRevB.55.7533}) the crossover temperature $T^*$ (labeled with coloured triangles and reported in the inset) grows significantly and the saturation value $\rho(T\rightarrow 0)$ decreases. 
Radiation-induced disorder (black and blue to green; from Wakeham \etal\cite{PhysRevB.94.035127}) only suppresses $\rho(T\rightarrow 0)$, while $T^*$ (labeled with coloured circles and reported in the inset) remains constant.
Also shown are results at ambient pressure from Katoh \etal \cite{Katoh199822} (dashed black line).
Differences between black curves (solid and dashed) demonstrate a strong sample dependence.
Inset: Dependence of $T^*$ on pressure (upper $x$-axis; red shaded triangles\cite{PhysRevB.55.7533}, grey diamonds from Campbell \etal\cite{PhysRevB.100.235133})
and disorder (measured in displacements per atom (DPA); lower $x$-axis; blue to green open circles\cite{PhysRevB.94.035127}).
}
\label{fig:exp}
\end{figure}

\paragraph{Introduction.}%
%
In Kondo insulators\cite{ki} the formation of bound-states between
quasi-localized $f$-states and conduction electrons leads to the opening of a narrow hybridization gap at the Fermi level. When this hybridization is coherent, the resistivity exhibits an activation-type behaviour. This semiconductor-like regime has as {\it upper bound} the Kondo lattice temperature, 
above which the local $f$-moments break free, inducing an insulator-to-metal crossover.
This Kondo effect has been exhaustively studied over the last decades\cite{Riseborough2000,Wirth2016,NGCS}. 
A more recent focus is the experimental observation of a {\it lower bound} to the semiconductor comportment, see \fref{fig:exp} for the example of Ce$_3$Bi$_4$Pt$_3$\cite{PhysRevB.55.7533,PhysRevB.94.035127,PhysRevB.100.235133}: Below an inflection temperature $T^*$ the resistivity levels off from exponential rise and enters a saturation regime---indicative of residual conduction.
Possible explanations include
classical exhaustion regimes (where {\it extrinsic} impurities pin the chemical potential) and
metallic {\it surface} states short-circuiting the gapped bulk. 
The latter can be an inevitable consequence of the non-trivial
nature of the insulating bulk found in topological Kondo insulators\cite{PhysRevLett.104.106408}.

Here, we develop a comprehensive 
perspective for residual conduction
 from {\it intrinsic bulk states with finite lifetimes}. 
We show that realistic many-body simulations capture the ill-understood resistivity in the (non-topological) Kondo insulator Ce$_3$Bi$_4$Pt$_3$. 
We then distill essential ingredients from a reductionist model, establish a microscopic understanding, 
and provide a 
phenomenological form of the resistivity with which experiments can be readily analysed.
Our theory is widely applicable to correlated narrow-gap semiconductors\cite{NGCS}: 
For mixed-valence SmB$_6$ we demonstrate that 
surface conduction
coexists with our mechanism for a residual bulk conductivity---%
providing a definitive interpretation of 
recent experiments\cite{Eo12638}.

 \begin{figure*}[!t!h]
    \centering
		\subfloat[simulated disorder at ambient pressure\label{subfig-1:dummy}]{\includegraphics[clip=true,trim=0 0 60 1,width=0.49\linewidth]{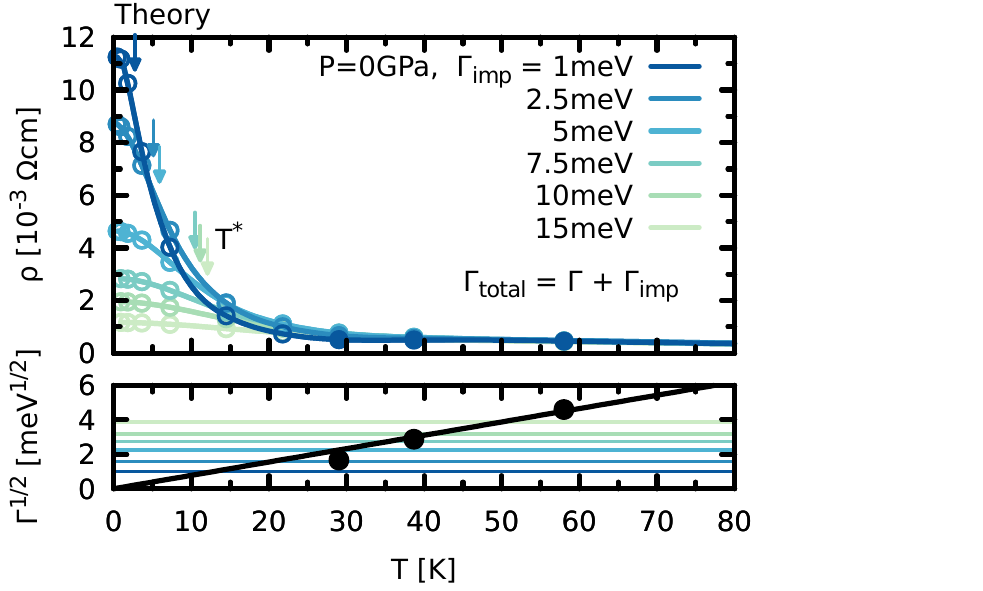}}  
    \subfloat[simulated pressure\label{subfig-2:dummy}]{\includegraphics[clip=true,trim=0 0 60 0,width=0.49\linewidth]{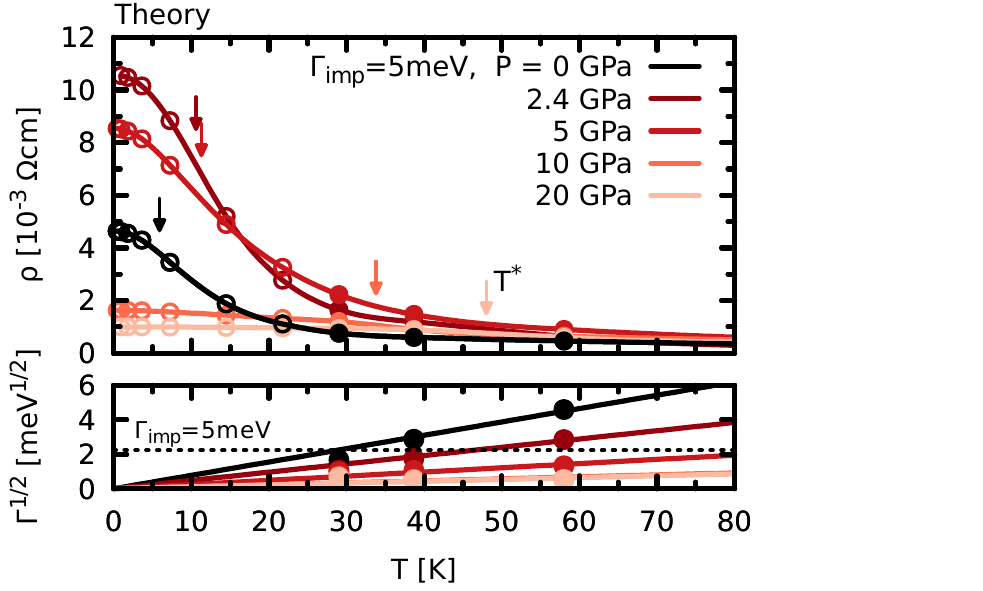}}
    \caption{{\bf Simulated resistivity of Ce$_3$Bi$_4$Pt$_3$.}
		Shown are results for $\rho(T)$ (a) at ambient pressure ($P=0$) for varying added impurity scattering $\Gamma_{imp}$ (shades blue to green)
		and (b) at different pressures $P$ (shades of red) for fixed impurity scattering $\Gamma_{imp}=5$meV.
		Filled circles in $\rho(T)$ indicate simulation temperatures; results shown by open circles have been obtained
		by extrapolating many-body lifetimes with a quadratic fit (see bottom panel and Method Section for details). 
		Vertical arrows indicate inflection points $T^*$ (reported also in \fref{fig:Tstar} top).
		The simulations mirror the experimental trends (cf.\ \fref{fig:exp}): $T^*$ varies significantly with pressure, but depends only weakly on disorder.
		The bottom panels display the square root of the simulated many-body scattering rate $\Gamma$ at the Fermi level (averaged over the Ce-4$f$ $J=5/2$ components);
		black and red-shaded lines are quadratic fits to the simulated points suggesting for all pressures the form
		$\Gamma=\gamma T^2$. $\gamma$ decreases notably under compression, indicative of weakening correlation effects.
		Horizontal lines (shades of blue and green in (a); dotted line in (b)) indicate the additional impurity-scattering rate $\Gamma_{imp}$. The total scattering rate is the sum of both: $\Gamma_{total}=\Gamma+\Gamma_{imp}$.
				}
				\label{fig:Kubo}
  \end{figure*}

\paragraph{The Kondo insulator Ce$_3$Bi$_4$Pt$_3$.}
Cubic intermetallic Ce$_3$Bi$_4$Pt$_3$ is a prototypical  Kondo insulator\cite{PhysRevB.42.6842,Riseborough2000,NGCS}:
Spectroscopic\cite{Takeda1999721,PhysRevLett.72.522} and
susceptibility\cite{PhysRevB.42.6842}
measurements (also in high magnetic field\cite{BOEBINGER1995227,Jaime2000})
are consistent with the Kondo scenario\cite{DONIACH1977231,jmt_CBP_arxiv}. 
While a topological bulk state has been envisaged\cite{Chang2017}, experiments argue against surface-dominated transport\cite{PhysRevB.94.035127}. 
Further constraints for a theory of resistivity saturation in Ce$_3$Bi$_4$Pt$_3$ come from, see \fref{fig:exp}: (i) pressure-dependent measurements that show a substantial increase in the crossover temperature $T^*$ accompanied by a decrease of
the saturation value $\rho(T\rightarrow 0)$\cite{PhysRevB.55.7533,PhysRevB.100.235133};
and (ii)
samples damaged by radiation in which residual conduction is successively
suppressed, while $T^*$ is unaffected\cite{PhysRevB.94.035127}.

\begin{figure}[!th]
{\includegraphics[width=\linewidth]{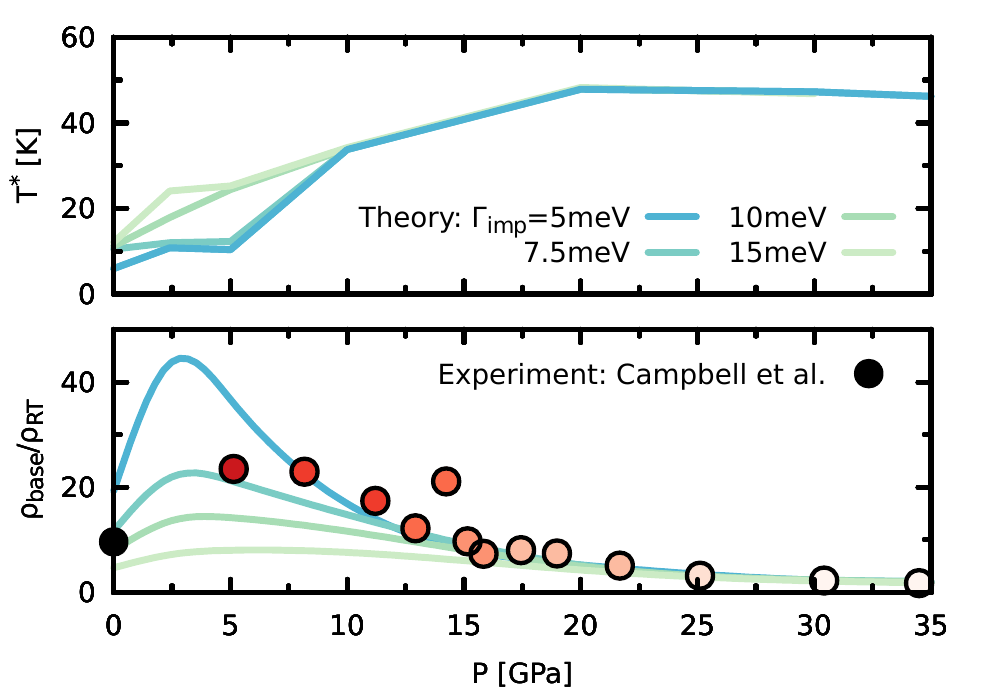}} 
\caption{{\bf Crossover temperature $T^*$ and resistivity ratio.} Top: Inflection temperatures $T^*$
as a function of pressure 
for different impurity-scatterings $\Gamma_{imp}$
(lines shaded blue to green). 
At low pressures, $T^*$ slightly depends on $\Gamma_{imp}$; above $P=5$GPa, the onset of saturation is insensitive to the magnitude of impurity scattering.
Bottom: 
Ratio of the resistivity at base temperature, $\rho_{base}$, and at
room temperature (RT), $\rho_{RT}$, for different impurity scatterings $\Gamma_{imp}$ (lines shaded blue to green), compared to experiment (circles; from Ref.~\onlinecite{PhysRevB.100.235133}).
In the simulation (experiment\cite{PhysRevB.100.235133})
the base and room temperatures were 1K and 290K (2K and 300K), respectively.
The overall trend of the resistivity ratio with pressure is independent of the strength of impurity scattering. 
We find best quantitative agreement with experiment for $\Gamma_{imp}\sim 7.5$meV.
}\label{fig:Tstar}
\end{figure}

\paragraph{Many-body simulations vs.\ Experiment.}
Using realistic many-body techniques, we simulate
the bulk response of Ce$_3$Bi$_4$Pt$_3$ under pressure.
In addition to renormalizations from electronic correlations---effective masses $m^*$ and scattering rates $\Gamma$, or lifetimes $\tau=\hbar/(2\Gamma)$---we mimic the effect of disorder\cite{sen2018fragility}
 with a temperature-independent $\Gamma_{imp}$, typical for impurity scattering\cite{coleman_2015}.
The resulting theoretical resistivities $\rho(T)$ are shown in \fref{fig:Kubo}
for varying (a) disorder 
and (b) pressure $P$. In all cases we identify
an inflection point $T^*$ below which a saturation regime emerges:
(a) At ambient pressure 
a growing  $\Gamma_{imp}$ 
causes $T^*$ to only marginally increase. The saturation value $\rho(T\rightarrow 0)$, however, is notably suppressed as lifetimes shorten---congruent with experiments (\fref{fig:exp}).
(b) Applying pressure 
boosts $T^*$ significantly until, see \fref{fig:Tstar} (top), it saturates between 20--30GPa---in qualitative agreement with experiment (\fref{fig:exp} inset).
The saturation limit $\rho(T\rightarrow 0)$ depends strongly
on pressure.
In experiments, the trend in $\rho(T\rightarrow 0)$ varies 
significantly between samples\cite{PhysRevB.55.7533} and setups\cite{PhysRevB.55.7533,PhysRevB.100.235133}. We therefore
follow Campbell \etal \cite{PhysRevB.100.235133} and reduce systematic errors by plotting
in \fref{fig:Tstar} (bottom) the ratio $\rho_{base}/\rho_{RT}$ of the simulated resistivity at the lowest (base) temperature ($T=1$K) and at room temperature (RT: $T=290$K).
Comparing to an experimental ratio
at similar temperatures\cite{PhysRevB.100.235133}, we see that
both {\it increase} from $P=0$ up to $\sim 3$--5GPa---the system becomes more insulating.
For higher pressures, however, the ratio {\it decreases} again---mirroring
the pressure-driven crossover to a bad insulator seen in \fref{fig:Kubo}(b).

The simulations thus contain the necessary ingredients  
for the observed resistivity saturation in Ce$_3$Bi$_4$Pt$_3$, including
its dependence on disorder and pressure.
Next, we characterize the saturation regime in more detail and elucidate its origin using a microscopic model.

 \begin{figure*}[!th]
    \centering
    \includegraphics[clip=true,trim=0 0 0 0,width=1.0\linewidth]{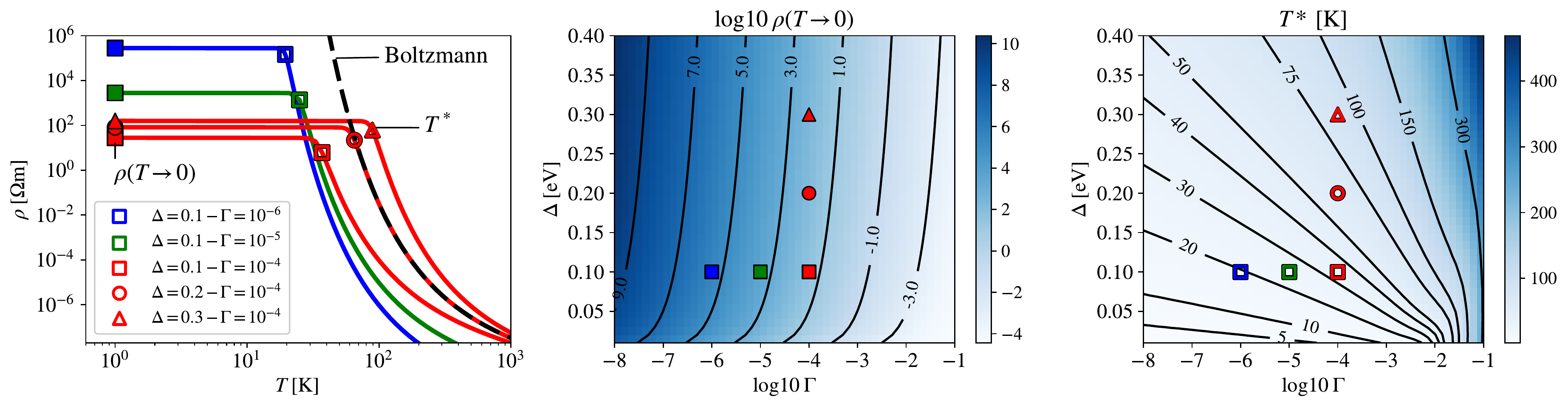}
    \parbox{\textwidth}{
    \qquad (a) resistivity $\rho(T)$\label{subfig-3:dummy}
    \hfill \qquad\qquad\qquad (b) residual resistivity $\rho(T\rightarrow 0)$\label{subfig-4:dummy}
    \hfill (c) crossover temperature $T^*$\qquad\label{subfig-5:dummy}
    }
    \caption{{\bf Prototypical resistivity in correlated narrow-gap semiconductors.}
    For the minimal two-band model (see text) the figure shows:
    (a) The resistivity (lines) as a function of temperature for different gaps $\Delta$ and scattering rates $\Gamma$ (both measured in eV). Closed symbols indicate the saturation limit $\rho(T\rightarrow 0)$; open symbols mark the inflection point $T^*$---the lower (upper) bound of the semi-classical (quantum) regime. 
    The semi-classical Boltzmann conductivity, \eref{Boltz}, is shown as dashed black line.
    (b) The saturation limit $\rho(T\rightarrow 0)$ (coloured map; on a log-scale) as a function of $\Delta$ and $\Gamma$; black lines are iso-curves for indicated values; colored symbols mark the choices of $\Delta$ and $\Gamma$ from panel (a). (c) Crossover temperature $T^*$ of the quantum regime. 
    The data shows that $T^*$ ($\rho(T\rightarrow 0)$) is dominantly controlled by $\Delta$ ($\Gamma$).
    Calculated for the three-dimensional half-filled two-band model, described in the main text, with hopping $t=0.25$eV, quasi-particle weight $Z=1$, and lattice constant $a=1$\AA.
    }
				\label{fig:LRT}
  \end{figure*}

\paragraph{Microscopic Theory.} 
We consider a half-filled two-band ($n=1,2$) model 
with hopping $t$ on the cubic lattice 
separated by a non-interacting gap $\Delta_0$:
$\epsilon^0_{\svek{k}n}=(-1)^n [2t\sum_{i=1,3}\cos(k_i) +  (6t + \Delta_0/2)]$.
We endow these dispersions with (i) a constant lifetime $\tau=\hbar/(2\Gamma_0)$ and (ii) a quasi-particle weight or mass enhancement $Z=m/m^*<1$.
The latter renormalizes the dispersion, $\epsilon_{\svek{k}n}=Z\epsilon^0_{\svek{k}n}$, yields the interacting gap $\Delta=Z\Delta_0$, and dresses the scattering rate $\Gamma=Z\Gamma_0$.
In the absence of particle-hole interactions, we can compute the linear-response conductivity of the model
exactly (see Method Section):
\begin{equation}
\sigma(T)=\frac{2\pi e^2}{\hbar V} \frac{Z^2}{4 \pi^3} \frac{\beta}{\Gamma}\sum_{\svek{k}n} v_{\svek{k}n}^2\left( \Re\Psi^\prime(z)- \frac{\beta\Gamma}{2 \pi}\Re\Psi^{\prime\prime}(z) \right)   
\label{LRT}
\end{equation}
with the inverse temperature $\beta=(k_B T)^{-1}$, the unit-cell volume $V$,
(derivatives of) the digamma function $\Psi(z)$ evaluated at
$z=\frac{1}{2}+\frac{\beta}{2\pi}\left(\Gamma+i\epsilon_{\svek{k}n}\right)$, and the Fermi velocities $v_{\svek{k}n}=1/\hbar\partial{\epsilon^0_{\svek{k}n}}/\partial \svek{k}$ in the Peierls approximation.
The above formula is rich in physics:
In the {\it coherent limit} $\Gamma\rightarrow 0$, \eref{LRT}
simplifies to the well-known Boltzmann expression in the constant relaxation-time approximation\cite{Ponce_2020}
\begin{equation}
\sigma(T)\stackrel{\Gamma\rightarrow 0}{=}\frac{e^2}{\hbar V}\frac{Z^2}{\Gamma}\sum_{\svek{k}n}v_{\svek{k}n}^2 \left(-\partial f/\partial\omega\right)_{\omega=\epsilon_{\svek{k}n}},
\label{Boltz}
\end{equation}
 with the Fermi function $f$---albeit with a renormalization $Z^2$ commonly not included.  
In this {\it semi-classical regime}, the conductivity is simply proportional to the lifetime 
 $\tau=\hbar/(2\Gamma)$. Then, for $k_BT\ll\Delta$, the resistivity has an activated
 form $\rho(T)\propto\exp(\Delta/(2k_BT))$
 that diverges for $T\rightarrow 0$.
 In fact, here, $\Delta$ is the only relevant energy scale: 
 As epitomized by Arrhenius-plot analyses, $\Delta$ single-handedly accounts 
 for the archetypal $\rho(T)$ of semiconductors in the Boltzmann regime. 
  To characterize the signatures of a {\it finite} scattering rate, $\Gamma>0$, we compute the resistivity $\rho(T)=1/\sigma(T)$ according to \eref{LRT},  
  see \fref{fig:LRT}.
 Akin to Ce$_3$Bi$_4$Pt$_3$, we see the emergence of a crossover temperature $T^*$ below which $\rho(T)$ tends towards saturation. In this {\it quantum regime} (QR), results deviate profoundly from the  Boltzmann limit.
In conventional semiconductors, deviations from activated behaviour typically occur when an {\it extrinsic in-gap density} pins the chemical potential. In our scenario, impurity states influence conduction merely by limiting the {\it lifetime of intrinsic} carriers.
Importantly, already minute scattering rates (mediated by impurities or other defects or couplings) lead to strong signatures at observable temperatures:
In \fref{fig:LRT}(b) and (c) we indicate, respectively, the saturation limit $\rho(T\rightarrow 0)$ and the characteristic temperature $T^*$ for the resistivities of panel (a).
In the relevant $\Gamma\ll\Delta$ regime, $T^*$ changes more rapidly with $\Delta$,
whereas $\rho(T\rightarrow 0)$ is more sensitive to changes in $\Gamma$---as in experiments and simulations for Ce$_3$Bi$_4$Pt$_3$ (see above).

We can give more precise analytical insight: At low temperatures, the  minimum (maximum) of conduction (valence) states dominates transport. For this leading contribution to \eref{LRT}, we neglect band-dispersions and 
consider two levels (2L) $\epsilon_n=(-1)^n\Delta/2$ ($n=1,2$) separated by a gap $\Delta$. 
Then, with $z=1/2+\beta/(2\pi)(\Gamma+i\Delta/2)$,
\begin{equation}
\sigma_{\hbox{\tiny 2L}}(T)\propto\frac{\beta}{\Gamma}\left[ \Re\Psi^\prime(z)- \frac{\beta\Gamma}{2 \pi}\Re\Psi^{\prime\prime}(z) \right].
\label{eq:two-level}
\end{equation}
As shown below, this phenomenological quantum conductivity---depending on the {\it two} energy scales $\Delta$ and $\Gamma$---is very useful for analysing experimental data.
A low temperature expansion of \eref{eq:two-level} to second order  yields
\begin{equation}
\sigma_{\hbox{\tiny 2L}}(T)\propto 
\frac{\Gamma^2}{(\Delta^2+4\Gamma^2)^2} \left( 1+\frac{8\pi^2}{3}\frac{5\Delta^2-4\Gamma^2}{(\Delta^2+4\Gamma^2)^2} \, (k_BT)^2 \right).
\label{eq:sigmaT2}
\end{equation}
resulting---for finite $\Gamma$---in the residual conductivity
\begin{equation}
    \sigma_{\hbox{\tiny 2L}}(T=0)\propto 
    \frac{\Gamma^2}{(\Delta^2+4\Gamma^2)^2}.
    \label{eq:sat}
\end{equation}
Unlike conduction by surface states in topological insulators, the quantum regime conductivity depends on the bulk values $\Delta$ and $\Gamma$.
Therefore, as a paramount distinction, residual conduction can be manipulated by pressure, while topological surface conduction is oblivious to it\cite{Cai2018}.
A direct consequence of \eref{eq:sat} is the existence of a temperature $T^*$
below which $\rho(T)$ departs from Boltzmann behaviour.
Using \eref{eq:sigmaT2}, we can estimate  the dependencies of $T^*$ via $\partial^2\rho(T)/\partial T^2=0$ (see Supplementary Fig.~S1):
\begin{equation}
k_BT^* = \frac{1}{\sqrt{10}\pi} \left(\frac{\Delta}{2} + \frac{11}{5} \frac{\Gamma^2}{\Delta} +\mathcal{O}(\Gamma^4) \right)
\label{eq:Tstar}
\end{equation}
For $\Gamma\ll\Delta$, $T^*$ is essentially controlled by $\Delta$---consistent with our numerical results and  available experiments.

The take-away message is this: If $\Gamma/\Delta$ is not vanishingly small, the lifetime of {\it intrinsic} charge carriers manifests as a
relevant energy scale. 
It introduces a coherence temperature $T^*$, delimits the applicability of Boltzmann theory from below, and leads to an algebraic saturation regime with residual conduction.

\begin{figure*}[!th]
 \subfloat[Pressure dependence]{\includegraphics[width=0.49\linewidth]{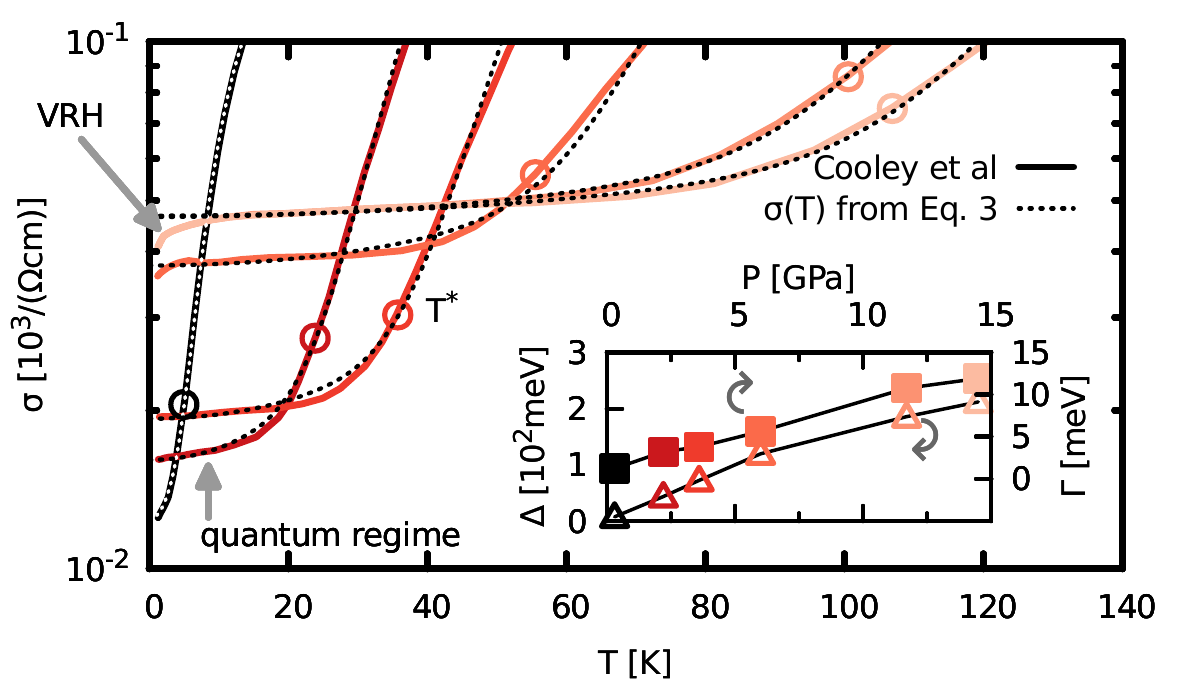}} 
 \subfloat[Influence of disorder]{\includegraphics[width=0.49\linewidth]{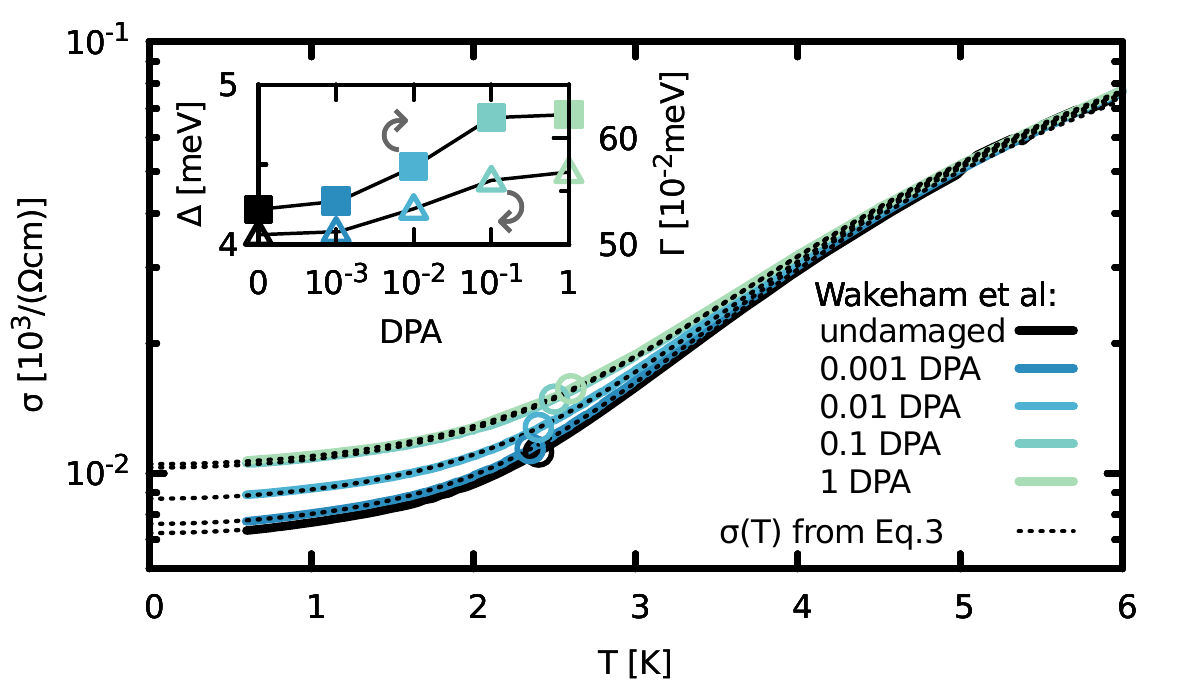}} %
\caption{{\bf Analysis of the quantum regime in Ce$_3$Bi$_4$Pt$_3$.}
We fit the conductivity $\sigma(T)=1/\rho(T)$ from (a)
 Cooley \etal \cite{PhysRevB.55.7533} (coloured lines correspond to pressures indicated in the inset) and (b)
 Wakeham \etal \cite{PhysRevB.94.035127} 
with the phenomenological quantum conductivity \eref{eq:two-level} (dashed  
lines).
The agreement 
is excellent: 
From a finite residual value for $T\rightarrow 0$,
the conductivity grows algebraically 
(see the quantum regime formula: \eref{eq:sigmaT2})
up to the crossover temperature $T^*$ (circles).
Above, higher powers in $T$ become relevant as $\sigma(T)$ enters the exponential (semi-classical) regime.
We analyse trends in the fit parameters (see insets):
(a) At $P=0$ we extract $\Delta\sim7.5$meV, which is largely enhanced for $P>0$; also $\Gamma$ increases with $P$.  
(b) 
We extract smaller $\Gamma$s and $\Delta$s for Wakeham \etal's sample,
owing to the overall smaller conductivity.
Consistent with the degree of radiation damage, $\Gamma$ increases with
growing displacements per atom (DPA).
The extracted $\Delta$ is congruent with activation-law fits above $T^*$\cite{PhysRevB.94.035127} and increases only marginally under radiation.
Note in (a),
for pressures $P>4$GPa and very low $T$, deviations from the quantum regime occur. There, as shown in Ref.~\onlinecite{PhysRevB.55.7533}, $\sigma(T)$ matches 3D variable-range hopping (VRH)  characteristics, $\propto\exp[(T/T_0)^{1/4}]$. %
Our fits used the approximation, \eref{eq:two-level}, that in particular neglects the momentum dependence of excitations. As illustrated in Supplementary Fig.~S2, this assumption generally leads to overestimation of, both, $\Delta$ and $\Gamma$.
Bare scattering rates $\Gamma_0$ (in analogy to $\Gamma_{imp}$ used in \fref{fig:Kubo}), can be obtained by multiplying $\Gamma$ with the mass enhancement $m^*/m=1/Z$.  
Because $m^*/m$ decreases with $P$ (see text), the pressure-driven increase in $\Gamma_0$ is smaller than for $\Gamma$. 
\label{fig:fit}}
\end{figure*}

\paragraph{Discussion.}
We now return to Ce$_3$Bi$_4$Pt$_3$. 
First, we analyze 
the experimental conductivity vis-\`a-vis 
the characteristic temperature profile of the quantum regime established above. 
Using the phenomenological quantum conductivity
\eref{eq:two-level}, we fit in \fref{fig:fit}
the data of (a) Cooley \etal\ \cite{PhysRevB.55.7533} and (b) Wakeham \etal \cite{PhysRevB.94.035127} and find near perfect agreement:
The microscopic Ansatz faithfully reproduces the experimental temperature dependence
for varying pressure and disorder
(see figure caption for details).
Adding this result to our realistic simulations discussed above,
we conclude that electronic scattering is the microscopic driver of the resistivity saturation.

\begin{figure}[!th]
\begin{center}
 \subfloat{\includegraphics[clip=true,trim=0 0 0 0 ,width=0.99\linewidth]{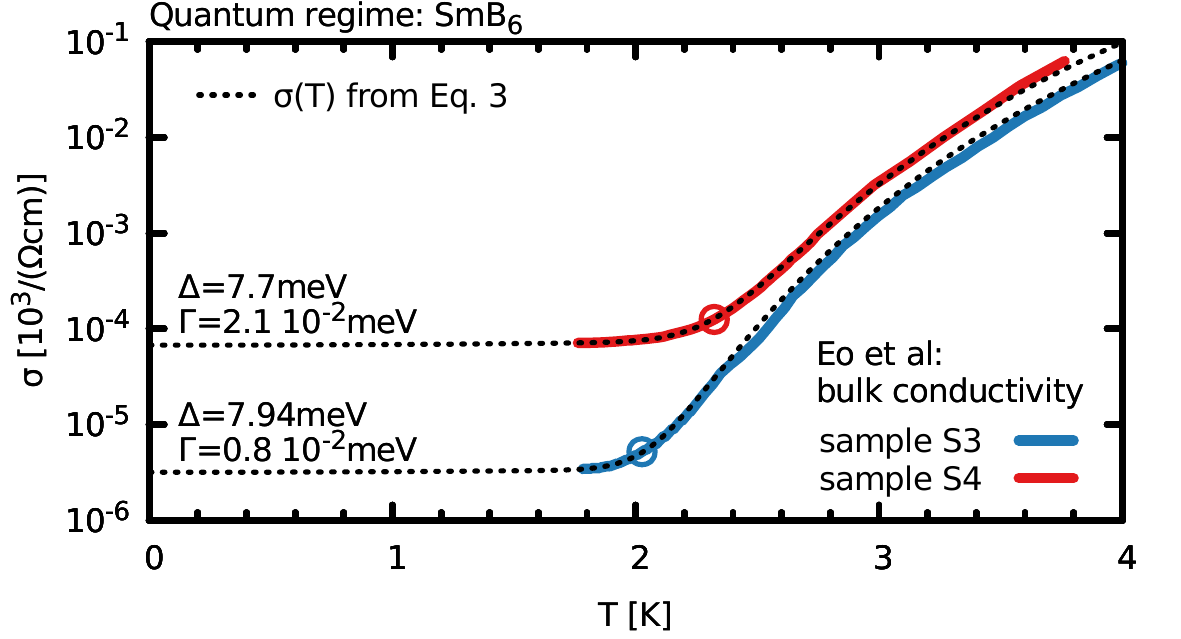}} 
 \end{center}
\caption{{\bf Bulk conductivity of SmB$_6$. } Low-temperature
transport in SmB$_6$ is dominated by surface contributions\cite{PhysRevB.88.180405,Kim2013,PhysRevLett.114.096601,Eo12638}.  
With an inverted resistance setup,
Eo \etal\ \cite{Eo12638} eliminated the latter and extracted the {\it bulk conductivity}, reproduced here for two samples, in order of growing off-stoichiometry: S3, S4.
Crucially, bulk conduction still saturates  below $T^*$ (inflection points in $\rho(T)$, circles).
We fit the experiment using  the phenomenological quantum conductivity
\eref{eq:two-level}.
The agreement is extraordinary, suggesting that the {\it bulk} 
resistivity in SmB$_6$ is lifetime-limited.
Gaps $\Delta$ are fixed to indicated values obtained in Ref.~\onlinecite{Eo12638} from activation laws, $\exp(\Delta/(2k_BT))$, above 3K. We extract a scattering rate that increases by more than two-fold for an off-stoichiometry that almost doubles from S3 to S4\cite{Eo12638}. 
%
%
\label{fig:SmB6}}
\end{figure}

Next, we briefly turn to
high pressures, where
  Campbell \etal\cite{PhysRevB.100.235133}
  found an insulator-to-metal crossover, cf.\ \fref{fig:Tstar} (bottom).
Indeed pressure-induced metallic phases are rather common for
correlated semiconductors, e.g., for
SmB$_6$\cite{PhysRevB.67.172406,ZHOU20171439}, 
CeRhSb\cite{doi:10.1143/JPSJ.65.27} and FeSi\cite{PhysRevB.100.155118}.
We consider three candidate mechanisms: 
Changes in (1) correlation effects, (2) the valence, (3) structural aspects.

(1) Our realistic calculations reveal that {\it pressure reduces  electronic correlations}: Effective masses shrink from $m^*/m\sim 10$ at $P=0$ (see also Ref.~\cite{jmt_CBP_arxiv}) to a mere $\sim2$ at $P=40$GPa. Also electron-electron scattering becomes less prevalent: 
The rate $\Gamma$ is---for all pressures---of the form $\gamma T^2$ (\fref{fig:Kubo} bottom panel), with $\gamma$ significantly decreasing with pressure. 
Reduced many-body renormalizations amplify the pressure enhancement of the non-interacting hybridization (see point (3), below), leading to larger gaps. In the absence of other factors, 
this is the canonical behaviour of Kondo insulators.

(2) Changes in the $f$-valency drives metal-insulator transitions, e.g., in
rare-earth monochalcogenides\cite{PhysRevB.87.115107}.
In Ce$_3$Bi$_4$Pt$_3$,  
we find pressure to not 
overturn the nominal $4f^1$ ($J=5/2$) ground state---excluding a dominantly valence-driven metallization (see Method Section).

(3) Instead, we unravel the non-monotonic transport to originate from two counter-acting {\it structural} trends within the confines of spacegroup $I\bar{4}3d$:
{\it Globally}, pressure shrinks the lattice, enhancing hybridization gaps.
While the atomic coordinates of Ce and Pt are dictated by symmetry, the {\it local} Bi position $(u,u,u)$ may vary.
Minimizing total energies, we find $u=0.088$ at $P=0$---in agreement with the experimental $u=0.086$\cite{PhysRevB.46.8067}---and predict a much larger $u=0.097$ at $P=40$GPa (see Supplementary Fig.~S3). Looking again at their data\cite{PhysRevB.100.235133}, Campbell \etal\ confirmed this trend\cite{pc_Campbell}. 
This seemingly minute modification drastically changes inter-atomic hybridizations: Instead of a monotonic increase (realized for $u=const$), a critical pressure emerges
above which the gap decreases%
\footnote{The critical pressure is $\sim 15$GPa in band-theory (see Supplementary Fig.~S3). The many-body simulated resistivity ratio in \fref{fig:Tstar} (bottom) drops earlier, consistent with experiment\cite{PhysRevB.100.235133}.}.
Hence, Ce$_3$Bi$_4$Pt$_3$ exhibits a peculiar high-pressure behaviour, not canonical for Kondo insulators in general.

\paragraph{Perspective.}
Our mechanism for resistivity saturation is relevant for other Kondo insulators. 
Indeed, also iso-structural
Ce$_3$Sb$_4$Pt$_3$ displays a $\rho(T)$\cite{PhysRevB.58.16057} consistent with our understanding: Different growth techniques (varying amounts of disorder) lead to largely different $\rho(T\rightarrow0)$ while $T^*$ changes little\cite{PhysRevB.58.16057}.
Ce$_3$Bi$_4$Pd$_3$, has recently been characterized as a semi-metal\cite{PhysRevLett.118.246601} or a Kondo insulator\cite{Kushwaha2019}. 
That the gap is next to non-existing\cite{PhysRevLett.118.246601,Kushwaha2019}
has been ascribed to spin-orbit\cite{PhysRevLett.118.246601} or Kondo\cite{jmt_radialKI} coupling effects.
Here, we conjecture that under compression a resistivity plateau develops in Ce$_3$Bi$_4$Pd$_3$. Future transport and susceptibility measurements should elucidate
whether pressurized Ce$_3$Bi$_4$Pd$_3$ mimics Ce$_3$Bi$_4$(Pd$_{1-x}$Pt$_x$)$_3$ for small $x$---(dis)favouring the (spin-orbit) Kondo scenario. %
Saturation tendencies have also been found in the Kondo insulators 
CeFe$_2$Al$_{10}$\cite{Muro2013} and pressurized CeRu$_4$Sn$_6$ \cite{PhysRevB.46.4250,STRYDOM2005293,ZhangCeRuSn}.

However, Kondo physics is not a prerequisite for our mechanism.
What makes these systems natural hosts for the quantum regime are their small gaps $\Delta\sim\mathcal{O}(\lesssim 50\hbox{meV})$. 
Correlation effects drive narrow gaps also in intermediate-valence insulators and $d$-electron intermetallics\cite{NGCS}, some of which do exhibit saturation regimes: SmB$_6$\cite{PhysRevB.88.180405,Kim2013,PhysRevLett.114.096601,Eo12638}, YbB$_{12}$\cite{Sato2019}  
or FeSb$_2$\cite{PhysRevB.88.245203}.  
How can we ascertain that our microscopic scenario is at work in any such compound? 
Salient signatures of the quantum regime provide guidance:
$T^*$ correlates with the bulk gap 
and the residual conductivity increases with shrinking lifetimes.
In mixed-valence SmB$_6$, however, the activation gap shrinks under pressure, while  $T^*$ is hardly affected\cite{PhysRevB.67.172406,ZHOU20171439} 
and added disorder at first increases the resistivity \cite{PhysRevB.94.035127}.
Also in CeRu$_4$Sn$_6$ single crystals, pressure significantly increases activation energies, while $T^*$ 
remains inert\cite{ZhangCeRuSn}.
These observations are incompatible with our scenario and suggest a different origin to dominate residual conduction%
\footnote{{\it Polycrystalline} CeRu$_4$Sn$_6$ samples
exhibit an additional inflection point in $\rho(T)$, which moves up under compression\cite{Sengupta_2012}.
We can hypothesize that the origin of this $T^*$--feature (that is consistent with our scenario) is induced by the larger disorder.
}.
Incidentally, for these two compounds
conducting surface states of proposedly topological character\cite{PhysRevX.7.011027,thunstrm2019topology} 
have been evidenced \cite{PhysRevB.94.035127,Eo12638}.

The situation of SmB$_6$ is, however, more complex:
Using a special measurement setup, Eo \etal\ \cite{Eo12638} were able to
disentangle surface and bulk contributions to conduction.
Crucially, the isolated {\it bulk conductivity} still exhibits a saturation regime---whose
temperature profile defies all previous scenarios\cite{Eo12638,PhysRevLett.121.026602}.
In \fref{fig:SmB6}, we demonstrate that the 
phenomenological
quantum conductivity \eref{eq:two-level} 
delivers an outstanding description of the experimental data---providing strong evidence that the bulk resistivity in SmB$_6$ is lifetime-limited.

We expect the characteristic temperature profile of the quantum regime to be ubiquitous for correlated narrow-gap semiconductors.
In fact, no matter how pure a sample can be prepared, the resistivity of no semiconductor will in practice diverge for $T\rightarrow 0$.
In the absence of other factors (or in combination with, see SmB$_6$), signatures of the presented physics should therefore manifest in any material with insulating ground-state---provided
low-temperature conduction is dominated by {\it intrinsic bulk} carriers.

\section{Methods} 
{\bf Realistic many-body electronic structure.}
We simulate Ce$_3$Bi$_4$Pt$_3$ at finite pressures using lattice constants from experimental fits to the third-order Birch-Murnaghan equation-of-state\cite{PhysRevB.100.235133} and relax
internal positions within density-functional theory as implemented in WIEN2k\cite{wien2k,wien2k2020} using the PBE functional.
Realistic dynamical mean-field theory (DMFT) calculations\cite{bible} are performed with the code of Haule \etal\cite{PhysRevB.81.195107}, including charge self-consistency, spin-orbit coupling, and
using a continuous-time quantum Monte-Carlo solver. Rotationally invariant interactions for the Ce-$4f$ shell  were parametrized by a Hubbard $U=5.5$eV and Hund's $J=0.68$eV  
At ambient pressure, this setup yielded excellent results 
for spectral and optical properties \cite{jmt_CBP_arxiv} (see also Ref.~\onlinecite{PhysRevLett.124.166403}).

The scattering rates $\Gamma$ displayed in \fref{fig:Kubo} (bottom) are obtained from the self-energy $\Sigma(\omega)$ by averaging over the $J=5/2$ components that dominate spectral weight near the Fermi level:
$\Gamma=-\Im\left\langle\Sigma(\omega=0)\right\rangle_{J=5/2}$. 
Finding that in all cases $\Gamma(T)\propto T^2$, 
we use this dependence to extrapolate the many-body scattering rate to temperatures beyond the reach of quantum Monte Carlo simulations.  In all cases, the crossover temperature $T^*$ occurs in a regime in which scattering is largely dominated by $\Gamma_{imp}$. Since we are interested in the system's properties around and below $T^*$, uncertainties in the extrapolation of $\Gamma(T)$ are hence negligible.

{\it Valency under pressure.}
Congruent with experiment\cite{PhysRevB.55.7533,PhysRevB.100.235133}
pressure decreases (increases) the  simulated Ce $4f$-occupation $n_f$ (valence $4-n_f$) from $n_f=1.05$ ($P=0$) to $0.96$ ($P=40$GPa)---while temperature has little influence\cite{PhysRevB.49.14708}.
$4f^0$ ($J=0$) admixtures augment with $P$---accounting for the larger valence. Yet, also $4f^2$ (and $4f^1$ with $J=7/2$) contributions grow---increasing the mixed valence character. Still, the probability of finding the system in a $4f^1$-state with $J=5/2$ merely decreases quantitatively from 80\% ($P=0$) to $\sim$55\% ($P=40$GPa)---excluding a valence-state transition. 

{\it Structural changes under pressure.}
The atomic position of Bi evolves with simulated pressure, see Supplementary Fig.~S3.
As a consequence the gap within band-theory grows only up to a critical pressure $P_c\sim 15$GPa. Beyond $P_c$, the centres of mass of conduction and valence states are still pushed apart, but a new dispersive conduction-band minimum emerges, that moves towards the Fermi level when increasing pressure further.

{\bf Transport properties.}
Resistivities for Ce$_3$Bi$_4$Pt$_3$ (Figs.~\ref{fig:Kubo}, \ref{fig:Tstar}) are simulated in linear-response with the full self-energy $\Sigma(\omega)$ as described in Ref.~\cite{jmt_fesi}, using dipole transition matrix elements\cite{AmbroschDraxl20061}. 
High-precision transport calculations that evaluate \eref{LRT} for the two-band model (\fref{fig:LRT}) are performed using 
{\texttt{\textsc{LinReTraCe}}}\cite{LRT}.
The realistic conductivities require a sizable impurity scattering rate $\Gamma_{imp}$ (broadening) for numerical stability. 
Comparing the resistivities in \fref{fig:exp} and \fref{fig:Kubo}, we see that simulated absolute values are on par with data from Katoh \etal\cite{Katoh199822}.
Experiments with better residual-resistance ratios (RRR)\cite{PhysRevB.55.7533,PhysRevB.94.035127,PhysRevB.100.235133}, have higher resistivities---suggesting that the broadening necessary in the simulations is too large for quantitatively mimicking high-quality samples.
{\texttt{\textsc{LinReTraCe}}}\cite{LRT}, on the other hand, yields numerically exact results for arbitrary scattering rates. 
The derivation of the central \eref{LRT} uses contour integration
techniques for the usual Kubo linear response for intra-band optical transitions with a static scattering rate, neglecting vertex corrections (following the DMFT spirit\cite{bible}).
Standard Fermi velocities, obtained by applying the Peierls approximation in the band-basis, $v_{\svek{k}n}=1/\hbar\partial{\epsilon^0_{\svek{k}n}}/\partial \svek{k}$, indeed only account for intra-band transitions. 
In a more general framework\cite{optic_prb} also inter-band transitions can be included 
within the Peierls philosophy. As far as the temperature dependence is concerned, 
inter- and intra-band contributions are very similar and the former have been omitted from
the model for clarity. For a discussion of intra- and inter-orbital transitions in the realistic simulations for Ce$_3$Bi$_4$Pt$_3$, see Supplementary Fig.~S4.
\eref{LRT} was first derived by one of us in Ref.~\cite{jmt_fesb2} in the context of thermoelectricity; here we study its merit for the low-temperature resistivity.
Crossover temperatures $T^*$ are extracted from the simulations of $\rho(T)$ at discrete temperature points using derivatives of a cubic spline interpolation.

In the Supplementary Material we include a minimal python script 
\verb=quantum_conductivity_fit.py=
with which experimental conductivities/resistivities can be analysed via \eref{eq:two-level}, see Supplementary Fig.~S5 for a screenshot.

{\bf Relation to other works.}
Our theory for resistivity saturation shares its key ingredient with the scenario of Shen and Fu \cite{PhysRevLett.121.026403} for the response in an external magnetic field:
There, {\it finite lifetimes} of Landau levels account for quantum oscillations in SmB$_6$\cite{Li1208,Tan287}.

{\bf Further technical remarks.}
The dependency of the resistivity on pressure and disorder are qualitatively captured in our realistic simulations, but the crossover temperature $T^*$ is underestimated. We can speculate on the origin of the quantitative discrepancy:
Pressure changes  how efficiently the bare Coulomb interaction is screened to yield the Hubbard $U$\cite{jmt_Wannier}.
The chosen interaction values provided quantitative results at ambient pressure\cite{jmt_CBP_arxiv}.
The trend towards metallicity under pressure (that dominate above $P\sim 3$--5GPa, see main text) may lead to more screening and, hence, to a smaller Hubbard $U$.
The latter will invariably lead to a larger charge gap and therefore (as explain in the main text) to a larger $T^*$.
Using projectors instead of Wannier orbitals to define locality in DMFT, a pressure-induced localization of orbitals\cite{jmt_radialKI,Chang_LSCO} that increases $U$ \cite{jmt_Wannier} further is not active here.

\bigskip

 \paragraph{Acknowledgements.}
%
The authors acknowledge discussions with D.\ J.\ Campbell and J.\ Paglione
and thank them for sharing unpublished information on atomic positions from their pressure experiments. 
The authors are grateful to F.\ Ronning, N.\ Wakeham and J.\ D.\  Thompson
for providing experimental raw data from Ref.~\cite{PhysRevB.94.035127}.
This work has been supported by the Austrian Science Fund (FWF)
through project {\texttt{\textsc{LinReTraCe}}} 
P~30213-N36. 
Calculations were partially performed on the Vienna Scientific Cluster (VSC).



%


\end{document}